\documentstyle[aps]{revtex}
\draft
\begin{document}
\title{PROTON MOMENTUM DISTRIBUTION IN NUCLEI BEYOND HELIUM-4}
\author{M.K. Gaidarov$^{1}$\footnote[0]{Permanent addresses:} \and
A.N. Antonov$^{2}$\footnote{Institute of Nuclear Research and Nuclear
Energy, Bulgarian Academy of Sciences, Sofia 1784, Bulgaria}  \and
G.S. Anagnostatos$^{2}$\footnote{Institute of Nuclear Physics, NCSR
"Demokritos", Aghia Paraskevi-Attiki, 15310 Greece}   \and
S.E. Massen$^{2}$\footnote{Department of Theoretical Physics,
University of Thessaloniki, GR-54006 Thessaloniki, Greece}  \and
 M.V. Stoitsov$^{1}$  \and \\ P.E. Hodgson$^{2}$}
\address{$^1$Institute for Nuclear Research and Nuclear Energy, Bulgarian
Academy of Sciences, Sofia 1784, Bulgaria}
\address{$^2$ Nuclear Physics
Laboratory, Department of Physics, University of Oxford, Oxford OX1-3RH,
U.K.}
\maketitle
\vspace{1cm}
\begin{abstract}
Proton momentum distributions of the $^{12}C$, $^{16}O$, $^{40}Ca$,
$^{56}Fe$ and $^{208}$Pb nuclei are calculated by a model using the
natural orbital representation and the experimental data for the momentum
distribution of the $^{4}He$ nucleus. The model allows realistic momentum
distributions to be obtained using only hole-state natural orbitals (or
mean-field single-particle wave functions as a good approximation to
them). To demonstrate the model two different sets of wave functions were
employed and the predictions were compared with the available empirical
data and other theoretical results.
\end{abstract}
\section{Introduction}
The systematic investigations of the nucleon momentum distributions in nuclei
extend the scope of the nuclear ground-state theory. Until the mid-seventies
more attention in the theory had been paid to the study of quantities such as
the binding energy and the nuclear density distribution $\rho (r)$. This is
related to the ability of the widely used Hartree-Fock theory to describe
successfully these quantities, which, however, are not very sensitive to the
dynamical short-range correlations. The experimental situation
in recent years concerning the interaction of particles with nuclei
at high energies, in particular the $(p,2p)$, $(e,e^{\prime}p)$ and
$(e,e^{\prime})$ reactions, the nuclear photoeffect, meson absorption by
nuclei, inclusive proton production in proton-nucleus collisions, and even some
phenomena at low energies such as giant multipole resonances, makes it possible
to study additional quantities. One of them is the nucleon momentum
distribution $n(k)$ \cite{1,2} which is specifically related to the processes
mentioned above. However, it has been shown \cite{3} that, in principle,
it is impossible to describe correctly both momentum and density distributions
simultaneously in the Hartree-Fock theory. The reason is that the
nucleon momentum distribution is
sensitive to short-range and tensor nucleon-nucleon correlations. It reflects
the peculiarities of the nucleon-nucleon forces at short distances which are
not included in the Hartree-Fock theory. This requires a correct simultaneous
description of both related distributions $\rho (r)$ and $n(k)$ in the
framework of nuclear correlation methods.

The main characteristic feature of the nucleon momentum distribution
obtained by various correlation methods [1, 2, 4--23] is the
existence of high-momentum components, for momenta $k>2$ fm$^{-1}$, due to the
presence of short-range and tensor nucleon correlations. This feature of $n(k)$
has been confirmed by the experimental data on inclusive and exclusive electron
scattering on nuclei (e.g. [1, 2, 24--27]). We emphasize also the fact that
theoretical results of various correlation methods \cite{9,18,19} as well as
experimental data for $n(k)$ obtained by the $y$-scaling analysis \cite{26}
confirm the conclusion \cite{5,28} that the high-momentum behaviour of the
nucleon momentum distribution
($n(k)$/$A$ at $k>2$ fm$^{-1}$) is similar for nuclei with mass
number $A$=2, 3, 4, 12, 16, 40, 56 and for nuclear matter (see \cite{2},
p.139).
More precisely, the high-momentum tails of $n(k)$ are almost the same for all
nuclei with
$A\geq4$ and thus $^{4}He$ is the lightest nuclear system that exhibits the
correlation
effects via the high-momentum components of the nucleon momentum distribution.
Since the magnitude of the high-momentum tail is proportional to the number of
particles, this effect is associated with the nuclear interior rather than
with the nuclear surface. This allows us in the present paper to suggest a
practical method to calculate the proton momentum distribution for nuclei
heavier than $^{4}He$ (e.g. $^{12}C$, $^{16}O$, $^{40}Ca$, $^{56}Fe$ and
$^{208}Pb$) from that one of $^{4}He$ which is already known from the
experimental data (or from calculations within correlation methods \cite{2}).
Here we should like to emphasize that though our method has some similarities
to that one suggested in \cite{16} (extended and developed in \cite{29,30,31}),
in contrast with the previous calculations, the correlation effects are
extracted from $^{4}He$ rather than from nuclear matter. We should like to
mention also that the experimental data for $n(k)$ in $^{4}He$ (which we use in
our calculations) as well as for other nuclei (which we use for a comparison)
are not directly measured but are obtained by means of the $y$-scaling analysis
\cite{26} relying on the assumption that the 1/$q$ expansion is valid. For this
reason we give an additional comparison of the data for $n(k)$ in $^{4}He$
\cite{26} with the theoretical calculations from \cite{19}.

In general, the knowledge of the momentum distribution for any nucleus is
important for calculations of cross-sections of various kinds of nuclear
reactions. It is known that reliable results for $n(k)$ in sophisticated
methods such as the $exp(S)$-method \cite{5}, the variational method with
state dependent correlations
\cite{20}, the generator coordinate method (with two generator coordinates
\cite{32}) and others are available only for light nuclei up to the $^{16}O$
nucleus. The local density approximation (LDA) which has been applied to derive
the spectral function of finite nuclei and to calculate $n(k)$ in $^{16}O$
\cite{29} was used as a basis to calculate $n(k)$ also in $^{40}Ca$ \cite{30}
and in nuclei with $A$=16, 40, 48, 90 and 208 \cite{31}.

The model suggested in this work uses the transparency of the single-particle
picture being within the framework of a given correlation method by means of
the natural orbital representation \cite{33}. The latter enables us to specify
in a natural way the high-momentum components in the momentum distribution
which are of the same magnitude for various nuclear systems. The theoretical
scheme of the method combines the mean-field predictions for the nucleon
momentum distribution which are expected to be realistic at small $k$ ($k\leq
2$ fm$^{-1}$) with the correlated part of the momentum distribution. In this
sense, the method in this work has a similarity with that one from \cite{29}
proposed for calculations of the spectral function $P({\bf k},E)$, whose energy
integral the nucleon momentum distribution is. We emphasize that in our work
this is done upon the common ground of the natural orbital representation. The
analyses of $n(k)$ performed in this work which use essentially the
correlations contained in $^{4}He$ nucleus (in contrast with the calculations
based on nuclear matter results already mentioned) can diminish the theoretical
uncertainties on $n(k)$ for medium-heavy nuclei.

\section{The model}

We start from the natural orbital representation \cite{33}, where the
proton momentum distribution normalized to unity is of the form \cite{2}:
\begin{equation}
n(k)=\frac{1}{4\pi Z} \sum_{nlj}{(2j+1)\lambda_{nlj}\left|
\widetilde{R}_{nlj}(k) \right|^{2}},
\label{1}
\end{equation}
where $\lambda_{nlj}$ is the natural occupation number for the state with
quantum numbers $(n,l,j)$ and
\begin{equation}
\sum_{nlj}(2j+1)\lambda_{nlj}=Z.
\label{new2}
\end{equation}
The radial part of the natural orbital in the momentum space
$\widetilde{R}_{nlj}(k)$ is related to the radial part of the natural orbital
in the coordinate space $\widetilde{R}_{nlj}(r)$ by
\begin{equation}
\widetilde{R}_{nlj}(k)=(2/\pi )^{1/2}(-i)^{l}\int_{0}^{\infty}
r^{2}j_{l}(kr)\widetilde{R}_{nlj}(r)dr,
\label{2}
\end{equation}
where $j_{l}(kr)$ is the spherical Bessel function of order $l$. We call
hole-state natural orbitals those natural orbitals for which the numbers
$\lambda_{nlj}$ are significantly larger than the remaining ones,
called particle-state natural orbitals \cite{34}.
It was shown by the Jastrow correlation method \cite{22} that
the high-momentum components of the total $n(k)$ caused by short-range
correlations are almost completely determined by the contributions of the
particle-state natural orbitals. This
fact, together with the approximate equality of the high-momentum tails
of $n(k)$ for all nuclei with $A\geq 4$, allows us to make the main
assumption of this work namely, that the particle-state contributions
to the momentum distribution are almost equal for all nuclei with $A\geq 4$.

Let us decompose the proton momentum distribution (\ref{1}) in two terms:
\begin{equation}
n(k)=n_{h}(k)+n_{p}(k),
\label{3}
\end{equation}
where the first term is the hole-state contribution
\begin{equation}
n_{h}(k)=\frac {1}{4\pi Z}\sum_{\alpha
(nlj)}^{\alpha_{F}} (2j+1)\lambda_{\alpha}\left|
\widetilde{R}_{\alpha}(k) \right|^{2},
\label{4}
\end{equation}
while the second one is the particle-state contribution
\begin{equation}
n_{p}(k)=\frac {1}{4\pi
Z}\sum_{\alpha_{F}}^{\infty}
(2j+1)\lambda_{\alpha}\left|
\widetilde{R}_{\alpha}(k) \right|^{2}.
\label{5}
\end{equation}
Using the assumed equality of the particle-state contributions $n_{p}(k)$ for
all nuclei, we obtain the following general relation of the correlated proton
momentum distribution of a nucleus $(A,Z)$
with that one of the $^{4}He$ nucleus:
\begin{equation}
n^{A,Z}(k) = N \left[ n^{^{4}He} (k) + \frac{1}{4\pi} \left(
\frac{1}{Z} \sum_{nlj}^{F_{A,Z}} (2j+1)
\lambda_{nlj}^{A,Z} \left| \widetilde{R}_{nlj}^{A,Z}
(k) \right|^{2} - \lambda_{1s_{1/2}}^{^{4}He} \left|
\widetilde{R}_{1s_{1/2}}^{^{4}He}(k) \right|^{2}
\right) \right],
\label{6}
\end{equation}
where
\begin{equation}
N = \left[ 1 + \frac{1}{Z} \sum_{nlj}^{F_{A,Z}} (2j+1) \lambda_{nlj}^{A,Z} -
\lambda_{1s_{1/2}}^{^{4}He}  \right]^{-1},
\label{7}
\end{equation}
and $F_{A,Z}$ is the Fermi level for the nucleus (A,Z).

Taking $^{40}Ca$ as an example the above expressions give:
\begin{eqnarray}
n^{^{40}Ca}(k) & = &\displaystyle  N \left[ n^{^{4}He } (k) +
\frac{1}{4\pi} \left( \frac{1}{10} \lambda_{1s_{1/2}}^{^{40}Ca } \left|
\widetilde{R}_{1s_{1/2}}^{^{40}Ca}(k) \right|^{2}
+ \frac{1}{5}\lambda_{1p_{3/2}}^{^{40}Ca} \left|
\widetilde{R}_{1p_{3/2}}^{^{40}Ca} (k) \right|^{2} \right. \right.
\nonumber \\
& + & \displaystyle \frac{1}{10}
\lambda_{1p_{1/2}}^{^{40}Ca} \left| \widetilde{R}_{1p_{1/2}}^{^{40}Ca}(k)
\right|^{2} + \frac{3}{10} \lambda_{1d_{5/2}}^{^{40}Ca} \left|
\widetilde{R}_{1d_{5/2}}^{^{40}Ca} (k) \right|^{2} +
\frac{1}{5} \lambda_{1d_{3/2}}^{^{40}Ca}
\left| \widetilde{R}_{1d_{3/2}}^{^{40}Ca} (k) \right|^{2} \\
& + & \displaystyle \left. \left. \frac{1}{10}
\lambda_{2s_{1/2}}^{^{40}Ca} \left| \widetilde{R}_{2s_{_{1/2}}}^{^{40}Ca} (k)
\right|^{2} - \lambda_{1s_{1/2}}^{^{4}He}
\left| \widetilde{R}_{1s_{1/2}}^{^{4}He} (k) \right|^{2}
\right) \right] \nonumber
\end{eqnarray}
with
\begin{equation}
N = \left[ 1 + \frac{1}{10} \lambda_{1s_{1/2}}^{^{40}Ca}  +
\frac{1}{5} \lambda_{1p_{3/2}}^{^{40}Ca} +
\frac{1}{10} \lambda_{1p_{1/2}}^{^{40}Ca} + \frac{3}{10}
\lambda_{1d_{5/2}}^{^{40}Ca} + \frac{1}{5} \lambda_{1d_{3/2}}^{^{40}Ca} +
\frac{1}{10} \lambda_{2s_{1/2}}^{^{40}Ca} -\lambda_{1s_{1/2}}^{^{4}He}
\right]^{-1}.
\label{11}
\end{equation}

As shown in \cite{22}, the hole-state natural orbitals are almost unaffected by
the short-range correlations and, therefore, the functions
$\widetilde{R}_{nlj}(k)$ in Eq. (\ref{6}) can be replaced by the corresponding
Hartree-Fock single-particle wave functions or by the shell-model
single-particle wave functions $R_{nlj}(k)$. The hole-state occupation numbers
$\lambda_{nlj}$ are close to unity within the Jastrow correlation method
\cite{22} and we can set them equal to unity with good approximation. The
properties of the hole-state natural orbitals and occupation numbers and the
decomposition of the proton momentum distribution in the hole- and
particle-state
contributions (Eqs.(4)-(6)) lead to a similarity of our model to that one
suggested for calculations of the spectral function in \cite{29}. In it the
mean-field predictions for the spectral function are combined with its
correlated part extracted from the nuclear matter calculations and recalculated
for finite nuclei within the local density approximation. In our model, the
correlated proton momentum distributions can be calculated for any nucleus by
means of the occupied shell-model wave functions and the proton momentum
distribution of the $^{4}He$ nucleus which is taken from \cite{26} and which
contains short-range correlation effects.

\section{Calculations and discussion}

In this work we calculate the proton momentum distribution for nuclei
$^{12}C$, $^{16}O$, $^{40}Ca$, $^{56}Fe$ and $^{208}Pb$. Empirical estimations
for $n(k)$ are available for nuclei $^{12}C$ and $^{56}Fe$ \cite{26}.

In our calculations of proton momentum distributions we use two types of MFA
single-particle wave functions: 1) single-particle wave functions obtained
within the Hartree-Fock method by using Skyrme effective forces and
2) multiharmonic oscillator single-particle wave functions
(with different values of the oscillator parameter for each state)
which lead to a simultaneous description of ground-state radii and binding
energies \cite{35,36}. In addition to \cite{36}, in our calculations the
multiharmonic oscillator s.p. wave functions are orthonormalized.
The values of all hole-state occupation probabilities $\lambda_{nlj}$ in
Eqs. (\ref{6}) and (\ref{7}) are set equal to unity. The empirical data
of $n(k)$ for $^{4}He$ are taken from \cite{26}. They are given in Table I. As
mentioned in the Introduction, the extraction of the data for $n(k)$ is
model-dependent. Due to this, we give in Fig.1 the comparison of the data for
$n(k)$ in $^{4}He$ from \cite{26} with the calculations within the variational
Monte Carlo method from \cite{19}. As can be seen from Fig.1, the agreement is
good and later we use in our calculations the data for $n(k)$ in $^{4}He$ from
\cite{26}.

The calculated proton momentum distributions for the nuclei examined are
given in Figures 2-6, respectively. They are compared with the available
data for $^{12}C$ and $^{56}Fe$ from \cite{26} and the proton momentum
distributions obtained in various theoretical methods, namely: for $^{12}C$
from \cite{22}, for $^{16}O$ from \cite{20,22,29,30,31}, for $^{40}Ca$ from
\cite{22,30} and for $^{208}Pb$ from \cite{31}. For the $^{12}C$ nucleus
the results for the proton momentum distribution using the s.p. wave
functions from the multiharmonic oscillator shell model but without
including correlations are given in Fig.7. Hence, the necessity of
accounting for correlations becomes apparent.

We have the following purposes within the practical method suggested in this
work for realistic calculations of the nucleon momentum distribution in light,
medium and heavy nuclei: 1) We like to show that the high-momentum tail of
the momentum distribution for any nucleus can be approximated by that for
$^{4}He$. We also check to what extent this approximation affects the central
part of the momentum distribution. Since the low-momentum components of $n(k)$
are determined mainly by the hole-state natural orbitals contribution,
the justification of the use of shell-model- or Hartree-Fock s.p. wave
functions instead of hole-state natural orbitals can be checked; 2) We
examine how well different s.p. shell-model wave functions can describe
also the middle part of the momentum distribution which bridges the
shell-model behaviour of the central part and the non-shell-model behaviour
of the tail of the momentum distribution; 3) We like to apply this method in
which correlation effects are extracted from $^{4}He$ to calculate $n(k)$ as
alternative one to the methods in which correlations are extracted from nuclear
matter and in this way, if possible, to diminish the theoretical uncertainties
on the momentum distribution for medium-heavy nuclei.

One can see from Figs.2 and 5 that the use of the single-particle wave
functions from the multiharmonic oscillator shell model leads to better
description of the experimental data for the central part of the momentum
distribution than the use of the Hartree-Fock single-particle wave functions.
In both cases the main deviations from the experimental data are for small
momenta ($k\leq 0.5$ fm$^{-1}$). They are larger in the case when Hartree-Fock
s.p. wave functions are used and this is a common feature of the results for
all nuclei considered. This is due to the well-known fact \cite{37} that the
Hartree-Fock method cannot give a realistic wave function for the $1s$ state in
the $^{4}He$ nucleus. Namely this function
($\widetilde{R}_{1s_{1/2}}^{^{4}He}(k)$) takes part in the expression for
$n(k)$ (Eq.(7)) in all nuclei. Both types of s.p. wave functions, however,
give similar results for the middle part as well as for the tail of
the momentum distribution in all cases considered.

The comparison of the results obtained by using of different mean-field
single-particle wave functions can be useful for the proper
choice of the latter in the applications of the model to practical calculations
of $n(k)$ in cases when the knowledge of this quantity is necessary.

Our numerical check shows that the results obtained by using values of the
occupation probabilities $\lambda_{nlj}$ coming either from the experiments or
from nucleon-nucleon correlation methods (e.g. Jastrow one \cite{22}) are
almost the same as those obtained with $\lambda_{nlj}$=1. As long as the
correlated values of $\lambda_{nlj}$ do not differ significantly from the
value $\lambda_{nlj}$=1, the improvement is not sizeable.

We emphasize that only hole-state occupation probabilities and wave functions
enter the main relationships of the model (Eqs.(7) and (8)). In this way the
suggested model can be easily applied to calculate momentum distributions in
nuclei taking into account the nucleon-nucleon correlation effects. Concerning
the particle-state contribution $n_{p}(k)$ (Eq.(6)) to the proton momentum
distribution (which is accounted for in the model by means of the term
$n^{^{4}He}(k)$ in Eq.(7)) we would like to mention that the decisive role for
the existence of the high-momentum components in $n_{p}(k)$ plays the form of
the particle-state natural orbitals $\tilde{R_{\alpha }}$ (which are strongly
localized in coordinate space) but not the particle-state occupation numbers
$\lambda_{\alpha }$ ($\alpha > \alpha_{F}$) \cite{1,2,13,22}.

As can be seen from Fig.3 the results for the proton momentum distribution
using the s.p. wave functions from the multiharmonic oscillator shell model are
in agreement (at least for $k>0.8$ fm$^{-1}$) with those obtained in the
variational Monte Carlo method \cite{20} and in the calculations based on the
local density approximation from \cite{29,30,31}. The same can be seen from the
comparison of our results for $^{40}Ca$ with those from \cite{30} given in
Fig.4. In our opinion, the calculations within the suggested model (with
correlation effects from $^{4}He$) and the similarity of the results with those
obtained in methods with correlations from nuclear matter can diminish in some
sense the theoretical uncertainties on $n(k)$. We would mention that some
differences remain between our results and those from \cite{31} for $^{208}Pb$
in the middle part of the momentum distribution (Fig.6).

It would be useful also to have estimations obtained extracting the correlation
effects from nuclei ligther than $^{4}He$, such as the deuteron or $^{3}He$. As
shown, however, in the variational correlation method (see Fig.5 in \cite{19}),
the high-momentum components of $n(k)$ for deuteron and $^{3}He$, although with
slopes similar with those of $n(k)$ for $^{4}He$ and nuclear matter, are quite
different from them. At the same time, the high-momentum components of $n(k)$
for $^{4}He$ and nuclear matter are almost the same. This is the reason to use
in our model correlation effects extracted from $^{4}He$ and to relate them
with the properties of the nuclear interior.

\section{Conclusions}

In the present paper a correlation model for calculating the proton momentum
distribution in nuclei with $A>4$ is proposed. The model combines the
mean-field part of the momentum distribution with its correlated part taken
from $^{4}He$ on one and the same footing using the natural orbital
representation. The estimation of the correlated part of $n(k)$ is based
on the well-known fact that the high-momentum components of the momentum
distribution normalized to unity (at $k\geq 2$ fm$^{-1}$) are nearly the
same for all nuclei with $A\geq4$. This fact, together
with the use of the natural orbital representation, gives the possibility
to obtain realistic momentum distributions in nuclei (including the regions
of small momenta, $k<2$ fm$^{-1}$, and of the intermediate momenta, $k\sim 2$
fm$^{-1}$) using only hole-state natural orbitals. The latter are replaced
to a good approximation by shell-model single-particle wave functions. Thus the
model gives a practical way for an easy calculation of the momentum
distribution for any nucleus. The numerical results in this work confirm to a
great extent the abilities of the suggested correlation model to give realistic
estimations for the proton momentum distribution in $^{12}C$ and $^{56}Fe$ and
to predict the behaviour of $n(k)$ in $^{16}O$, $^{40}Ca$ and $^{208}Pb$
nuclei. They are in agreement with the results for the proton momentum
distribution in $^{16}O$ and $^{40}Ca$ obtained within other theoretical
methods in which the correlation effects are incorporated using nuclear matter
results and with some empirical data for $^{12}C$ and $^{56}Fe$ obtained using
the $y$-scaling method. The knowledge of the realistic proton momentum
distributions obtained in this work would allow us to describe in a similar way
as it is done in \cite{38} quantities which are directly measurable in
processes of particle scattering by nuclei.

\acknowledgments

Three of the authors (A.N.A, G.S.A and S.E.M) are grateful to the
Nuclear Physics Laboratory of the University of Oxford for the kind
hospitality.
The author A.N.A. is grateful to the Royal Society and the Bulgarian Academy
of Sciences for the support during his visit in the University of Oxford.
The authors M.K.G., A.N.A. and M.V.S. would like to thank the Bulgarian
National Science Foundation for partial finantial support under the Contracts
Nrs.$\Phi$--32 and $\Phi$--406. Finally,
the authors G.S.A. and S.E.M. would like also to thank their home institutions
for granting their sabbatical leaves.

\newpage
\noindent
Figure 1. Proton momentum distribution $n(k)$ versus $k$ of $^{4}He$. The solid
triangles represent the data from \protect\cite{26}. The solid line is the
result from \protect\cite{19}. The normalization is:
$\displaystyle \int n(k)d^{3}{\bf k}=1$.
\vspace{0.7cm}\\
Figure 2. Proton momentum distribution $n(k)$ versus $k$ of $^{12}C$.
Calculations by using single-particle wave functions from the multiharmonic
oscillator shell model \protect\cite{36} are presented by solid line
and those by using
Hartree-Fock single-particle wave functions by long-dashed line. The
short-dashed line is $n(k)$ calculated in the Jastrow correlation method
\protect\cite{22}. The solid triangles represent the data
from \protect\cite{26}. The normalization is as in Fig.1.
\vspace{0.7cm}\\
Figure 3. Proton momentum distribution $n(k)$ versus $k$ of $^{16}O$.
The lines 1 and 2 are the results of the present work using Hartree-Fock and
multiharmonic oscillator shell model \protect\cite{36} single-particle wave
functions, respectively. The lines 3, 4, 5, 6 and 7 are the results from
\protect\cite{22}, \protect\cite{30}, \protect\cite{20}, \protect\cite{31}
and \protect\cite{29}, respectively. The normalization is as in Fig.1.
\vspace{0.7cm}\\
Figure 4. Proton momentum distribution $n(k)$ versus $k$ of $^{40}Ca$.
The solid and long-dashed lines are as in Figure 2. The dotted line is $n(k)$
calculated in \protect\cite{22}. The short-dashed line is the result from
\protect\cite{30}. The normalization is as in Fig.1.
\vspace{0.7cm}\\
Figure 5. Proton momentum distribution $n(k)$ versus $k$ of $^{56}Fe$.
Calculations by using single-particle wave functions from the multiharmonic
oscillator shell model \protect\cite{36} are presented by solid line and
those by using Hartree-Fock single-particle wave functions by long-dashed
line. The solid triangles represent the empirical data from
\protect\cite{26}. The normalization is as in Fig.1.
\vspace{0.7cm}\\
Figure 6. Proton momentum distribution $n(k)$ versus $k$ of $^{208}Pb$.
Calculations by using single-particle wave functions from the multiharmonic
oscillator shell model \protect\cite{36} are presented by solid line and
those by using Hartree-Fock single-particle wave functions by long-dashed
line. The results from \protect\cite{31} are given by short-dashed line.
The normalization is as in Fig.1.
\vspace{0.7cm}\\
Figure 7. Proton momentum distribution $n(k)$ versus $k$ of $^{12}C$.
Calculations by using single-particle wave functions from the multiharmonic
oscillator shell model \protect\cite{36} without including correlations are
presented by solid line. The solid triangles represent the empirical data
from \protect\cite{26}. The normalization is as in Fig.1.

\newpage
\begin{table}
\caption{Experimental values of the proton momentum distribution in
$^{4}He$ \protect\cite{26}. The normalization is:
$\displaystyle \int n(k)d^{3}{\bf k}=1$.}
\vspace{0.2cm}
\begin{tabular}{cccc}
k & n(k)  &  k & n(k)        \\
MeV/c & $fm^{3}$  & MeV/c & $fm^{3}$\\
\cline{1-2} \cline{3-4}
   50    &    0.757916   & 350    &    0.001065  \\
  100    &    0.331900   & 400    &    0.000749  \\
  150    &    0.119684   & 450    &    0.000615  \\
  200    &    0.039692   & 500    &    0.000548  \\
  250    &    0.011019   & 550    &    0.000380  \\
  300    &    0.002855 &  &\\
\end{tabular}
\end{table}
\end{document}